\documentstyle[preprint,aps]{revtex}

\tightenlines

\def\x{\times}
\def\P{{\cal P}}
\def\Q{{\cal Q}}
\def\m{{\cal M}}

\preprint{hep-th/9712224}

\draft

\title{Black hole entropy as T-duality invariant}

\author{ Kenji Suzuki }
\address{
\it Department of physics, Tokyo Institute of Technology \\
\it Oh-okayama, Meguro, Tokyo 152, Japan \\
\it ks@th.phys.titech.ac.jp
}

\begin{document}

\maketitle

\begin{abstract}
We study the Euler numbers and 
the entropies of the non-extremal intersecting 
D-branes in ten-dimensions.
We use the surface gravity to constrain the compactification radii. 
We correctly obtain the integer valued Euler numbers for these radii.
Moreover, the entropies are found to be invariant 
under the T-duality transformation. 
In the extremal limit, we obtain the finite entropies only for 
two intersecting D-branes. We observe that these entropies are 
proportional to 
the product of the charges of each D-brane. 
We further study the entropies of the boosted metrics. 
We find that their entropies can be interpreted 
in term of the microscopic states of D-branes. 
\end{abstract}



\section{Introduction}
The Bekenstein-Hawking entropies of the BPS saturated 
black holes have been studied in four-, five- 
and ten-dimensions from the nonperturbative aspects of string theory
\cite{CY,CT,SV,HS,MS,Tsey,GKT,BB}. 
The black holes in ten-dimensions are constructed 
by the intersecting D-branes. 
The black holes are dual to each other 
under the T-duality transformation. 
The four- and five-dimensional black holes 
are obtained by dimensional reduction 
from the ten-dimensional black holes. 
The method to obtain the non-extremal black holes 
from the extremal black holes are studied. 
Their entropies are interpreted in terms 
of the microscopic states. 
The effect of the graybody factor to the Hawking-radiate 
blackbody radiation is also derived from the D-branes. 

On the other hand, the Euler numbers of the black holes are 
investigated in four-dimensions~\cite{HHR,GK,LP}.
The Euler number $\chi$ of an $n$-dimensional manifold 
is the sum of the Betti numbers $B_p$: 
\begin{eqnarray*}
 \chi = \sum_{p=0}^{n}(-1)^p B_p.
\end{eqnarray*}
The Betti numbers are integers. 
Therefore the Euler numbers are also integers. 
We can also calculate the Euler numbers by  
using the $n$-dimensional Gauss-Bonnet action, 
which integrates the curvatures $R_{ij}$ 
of the black holes. 
The $n$-dimensional Gauss Bonnet action is 
\begin{eqnarray*}
 \chi &=& \frac{1}{(4\pi)^{n} n!} \int dx^n
 \epsilon^{i_1i_2 \cdots i_n}
  R_{i_1i_2}R_{i_3i_4} \cdots R_{i_{n-1}i_n} .
\end{eqnarray*}
The black holes in four-dimensions have the period 
in Euclidean time coordinate. 
The period is defined by the inverse of the surface gravity. 
By using this relation, we can obtain the integer valued Euler numbers. 

In the same way as in four-dimensions, 
we can calculate the Euler numbers of the black holes 
in ten-dimensions using the ten-dimensional Gauss-Bonnet action. 
The Euler numbers of the black holes in ten-dimensions 
have not been studied so far. 
We consider the Euler numbers 
of the black holes which are realized by the intersecting D-branes 
in ten-dimensions. 
The black holes of the intersecting D-branes in ten-dimensions 
also have the periods with respect to the compactification radii. 
The radii are usually taken to be arbitrary. 
For the generic radii, 
the Euler numbers calculated 
by the ten-dimensional Gauss-Bonnet actions are non-integers. 
However the Euler numbers must be integers because of 
their definition. 
To avoid these difficulties, we propose a way to constrain 
the compactification radii. 

We reflect on the necessity 
to fix the values of the compactfication radii 
in the physical reasoning. 
We treat the black holes in ten-dimensions. 
These have six compactified directions. 
The entropies of the black holes in ten-dimensions 
are expressed by the compactification radii 
and the charges.
We rewrite the entropies 
using the quantized charges 
and the Newton's constant, 
then the entropies are irrespective 
of the compactification radii. 
Therefore it seems not to be necessary to fix 
the value of the compactification radii. 
However, the four-dimentional Newton's constant $G_{(4)}$
depends on the value of the compactified radii,  
\begin{eqnarray*}
 G_{(4)} &=& G_{(10)}/L_1L_2L_3L_4L_5L_6 ,
\end{eqnarray*}
where $L_i$ are the compactification radii. 
$G_{(10)}$ is the ten-dimensional Newton's constant,
\begin{eqnarray*}
 G_{(10)} = 8\pi^6 g^2 
\end{eqnarray*}
with $\alpha' = 1$. 
We find that 
the compactified radii depend on 
the radius of the horizon ($\mu$) and the charges, 
using the method of the surface gravities 
of the compactified directions,
discussed in the section 3.
Then the four-dimensional Newton's constant 
is also rewritten by $\mu$ and the charges. 
We are interested in the behavior of 
the four-dimensional Newton's constant 
in the BPS limit, $\mu \to 0$.
As a result, we obtain that 
the Newton's constant vanishes in the BPS limit. 

The another physical reasoning that we need to fix 
the compactification radii 
is to avoid the singular effects 
of the horizon in the compactified directions. 
We recall the way 
to define the period in the time direction.
If no singular effects of the horizon exist in $t-r$ directions, 
then the topology of these directions are ${\bf R^2}$, 
and the Euler number of these directions is 1. 
Using the Gauss-Bonnet theorem,
we find that 
the Euclidean time coordinate 
have the period which is 
the inverse of the surface gravity. 
Therefore we need to take this period 
in the time direction 
to avoid the singular effects of the horizon.
Similarly, we need to fix 
the compactification radii to avoid the singular effects 
of the horizon in the compactified directions.
We find that 
the compactification radii are the inverses of 
the surface gravity in the compactified directions 
using the Gauss-Bonnet theorem. 
We discuss this point in the section 3 .


The purpose of this paper is to constrain 
the compactification radii 
in order to obtain the integer valued Euler numbers 
for the non-extremal black holes of the intersecting D-branes 
in ten-dimensions. 
We further study the microscopic interpretations of their 
entropies. 

We consider the black holes of the intersecting D-branes 
which have some compactification radii 
in ten-dimensions~\cite{CT,Tsey,GKT}. 
We assume that the metrics only depend on 
$r = \sqrt{x^2+y^2+z^2}$.
In the four-dimensions, the black holes have the period 
in Euclidean time coordinate. 
The period is defined by the inverse of the surface gravity.
We extend this construction to constrain the compactified radii 
in ten-dimensions. 
We use the inverse of the surface gravities 
in the compactified 
directions to constrain these radii. 
As a result, we are able to obtain the integer Euler numbers. 
We recall that the temperatures are defined by the surface gravities 
in the time coordinates. 
The relation between the radii and the surface gravities 
generalizes the temperatures of the black holes. 

Using these results, 
we further obtain the entropies of the non-extremal
intersecting D-branes.
These entropies are invariant under T-duality transformation. 
In the BPS limit, 
we obtain the finite and non-vanishing entropies 
only for two intersecting D-branes. 
We observe that they can be interpreted as the product 
of charges of each D-brane. 
We then study the entropies of the boosted metrics. 
We obtain the famous relation 
between the entropies and the quantized D-brane charges 
and the internal momenta of the intersecting D-branes. 
These relations are the same as 
that of the microscopic D-brane picture~\cite{SV,HS,MS}. 

The organization of this paper is as follows. 
In section 2, we review the way to construct the intersecting 
D-branes of the type IIA and IIB superstrings.
We explain how to obtain the non-extremal black holes 
from the extremal black holes. 
In addition we define the boost transformation of the metrics.
In section 3, we consider the surface gravities 
in the compactified directions 
to constrain the compactification radii. 
As a result, we obtain the Euler numbers which are integers. 
In section 4, we calculate the entropies of the non-extremal 
intersecting D-branes and study them in the BPS limit.
We further discuss the interpretation of these entropies 
in terms of the microscopic states. 

\section{Non-extremal Intersecting D-branes in ten-dimensions}
In this section we review the way 
to construct the intersecting D-branes~\cite{Tsey}.
We then explain how to obtain non-extremal black holes 
from the extremal black holes and to obtain the boosted metrics. 
We treat the intersecting D-branes in ten-dimensions with the
three-dimensional transverse directions with respect to the D-branes. 
This means that 
the harmonic functions of these metrics only depend on 
$r = \sqrt{x^2+y^2+z^2}$.

We introduce the metrics and the field strengths in association with 
the D-branes. 
The metrics of the D$n$-branes wrapped on the
$i_1, i_2, \cdots i_n$- directions are 
\begin{eqnarray}
   ds^2_{10}  = H^{1/2} 
    \big[ H^{-1}(-dt^2 + dx_{i_1}^2 + \cdots + dx_{i_n}^2 )
          + dr^2 + r^2 d\Omega^2_{10-n-2} \big] .
\end{eqnarray}
$H$ is the harmonic function. 
The field strengths are 
\begin{eqnarray*}
  {\cal F} = \left \{
\begin{array}{@{\,}ll}
 dt \wedge dH^{-1} \wedge dx_{i_1} \wedge 
 \cdots \wedge dx_{i_n} 
  & (n\le 3; \ electric) \\ 
 \ast(dt \wedge dH \wedge dx_{i_1} \wedge 
 \cdots \wedge dx_{i_n})  
  & (n\ge 3; \ magnetic) , 
\end{array}
\right.  
\end{eqnarray*}
where $\ast$ is the Hodge dual in ten-dimensions.
In $n=3$, the field strength is self-dual. 
Then they have the field strengths 
both of electric and magnetic nature. 
For the intersecting D-branes, ${\cal F}$ are 
the sum of the field strengths of each D-brane. 
We obtain the metrics of the type IIA string theory 
from the intersecting M-branes by dimensional reductions.
The intersecting M-branes with the three-dimensional 
transverse directions with respect to the M-branes 
are $(2,2,5,5)_M$, $(5,5,5)_M$, $(2,5,5)_M$, 
$(2,2,5)_M$, and $(5,5)_M$, where $(5,5)_M$ denote two 
intersecting M5-branes, and so on~\cite{CT,Tsey,GKT}.

We subsequently review the metrics of three intersecting D-branes.
In ten-dimensions, we can obtain the (4,4,4) metric 
from the intersecting M-branes $(5,5,5)_M$ 
upon dimensional reduction along a common direction 
of three intersecting M5-branes :
\begin{eqnarray}
   ds^2_{10}  &=& (H_1 H_2 H_3)^{1/2} 
    \big[ - H_1^{-1} H_2^{-1} H_3^{-1}dt^2 \nonumber \\
 &&   + H_1^{-1} H_2^{-1} (dx_1^2 + dx_2^2) 
      + H_1^{-1} H_3^{-1} (dx_3^2 + dx_4^2) \nonumber \\
 &&   + H_2^{-1} H_3^{-1} (dx_5^2 + dx_6^2)
      + dr^2 + r^2 d\Omega^2_2 \big] .
\end{eqnarray}
The field strength ${{\cal F}_4}_{(4,4,4)}$ 
and the dilaton $\phi$ are  
\begin{eqnarray}
 {{\cal F}_4}_{(4,4,4)}
 &=& (\partial_r H_1 \ d\theta \wedge d\phi \wedge dx_6 \wedge dx_7 
  \nonumber \\
 && + \partial_r H_2 \ d\theta \wedge d\phi \wedge dx_4 \wedge dx_5 
    + \partial_r H_3 \ d\theta \wedge d\phi \wedge dx_1 \wedge dx_2), \\ 
 e^{-2\phi} &=& (H_1H_2H_3)^{1/2}, 
\end{eqnarray}
where $H_i = 1+ \frac{Q_i}{r}$. We denote this metric 
as the following~\cite{BB} : 
\begin{eqnarray}
 (4,4,4) &=& \mbox{ {\scriptsize
        $\left\{ \begin{array}{c|cccccccccc}
     \x & \x & \x & \x & \x & -  & -  & -  & -  & - & \\
     \x & \x & \x & -  & -  & \x & \x & -  & -  & - & \\
     \x & -  & -  & \x & \x & \x & \x & -  & -  & - & .
               \end{array} \right. $} } \label{444}
\end{eqnarray}

Similarly we can obtain the $(2,4,4)$ metric, 
from $(2,5,5)_M$ upon dimensional reduction 
along a direction of two intersecting M5-branes, 
\begin{eqnarray}
 (2,4,4) &=& \mbox{ {\scriptsize
         $\left\{ \begin{array}{c|cccccccccc}
       \x & \x & \x & -  & -  & -  & -  & -  & -  & - & \\
       \x & \x & -  & \x & \x & \x & -  & -  & -  & - & \\
       \x & -  & \x & \x & \x & -  & \x & -  & -  & - & .
               \end{array} \right. $} }   
\end{eqnarray}
The field strength ${{\cal F}_4}_{(2,4,4)}$ 
and the dilaton $\phi$ are  
\begin{eqnarray}
 {{\cal F}_4}_{(2,4,4)}
 &=& ( dt \wedge \partial_r H_1^{-1} \ dr \wedge dx_1 \wedge dx_2 
  \nonumber \\
 && + \partial_r H_2 \ d\theta \wedge d\phi \wedge dx_2 \wedge dx_6 
    + \partial_r H_3 \ d\theta \wedge d\phi \wedge dx_1 \wedge dx_5), \\
 e^{-2\phi} &=& (H_1^{-1}H_2H_3)^{1/2}. 
\end{eqnarray}

Along a direction of the M5-brane, we can obtain the $(2,2,4)$ metric, 
\begin{eqnarray}
 (2,2,4) &=& \mbox{ {\scriptsize
         $\left\{ \begin{array}{c|cccccccccc}
      \x & \x & \x & -  & -  & -  & -  & -  & -  & - & \\
      \x & -  & -  & \x & \x & -  & -  & -  & -  & - & \\
      \x & \x & -  & \x & -  & \x & \x & -  & -  & - & .
               \end{array} \right. $} }  
\end{eqnarray}
The field strength ${{\cal F}_4}_{(2,2,4)}$ 
and the dilaton $\phi$ are 
\begin{eqnarray}
 {{\cal F}_4}_{(2,2,4)}
 &=& (dt \wedge \partial_r H_1^{-1} \ dr \wedge dx_1 \wedge dx_2 
    \nonumber \\  
 && +  dt \wedge \partial_r H_2^{-1} \ dr \wedge dx_3 \wedge dx_4 
    + \partial_r H_3 \ d\theta \wedge d\phi \wedge dx_2 \wedge dx_4), \\
 e^{-2\phi} &=& (H_1H_2H_3^{-1})^{-1/2}.
\end{eqnarray} 
In these constructions, we have regarded the D2-branes as electric 
and the D4-branes as magnetic objects.

Moreover, we can obtain the remaining metrics 
by T-duality transformation from these metrics 
in the $i$-th direction~\cite{GKT} : 
\begin{eqnarray*}
 g'_{ii} = 1/g_{ii}, \quad 
 \exp(-2\phi') = g_{ii} \exp(-2\phi) .
\end{eqnarray*}
In our notation, the T-duality transformation 
is realized by interchanging 
$\x$ and $-$. For example, in the three intersecting D-branes case,  
\begin{eqnarray}
 \mbox{ {\scriptsize
        $\left\{ \begin{array}{cc}
         \x & \\
         \x & \\
         -  & 
 \end{array} \right. $} } \leftrightarrow 
 \mbox{ {\scriptsize
        $\left\{ \begin{array}{cc}
         -  &  \\
         -  &  \\
         \x & .
                 \end{array} \right. $} }
\end{eqnarray}
If we apply the T-duality transformation 
on the metric $(4,4,4)$ (in (\ref{444}))
in the first direction, this metric is changed to $(3,3,5)$. 
Similarly, we can obtain the following metrics :
\begin{eqnarray*}
 && (4,4,4) \leftrightarrow
    (3,3,5) \leftrightarrow
    (2,4,4) \ or \ (2,2,6) \\
 &&  \leftrightarrow 
    (3,3,3) \ or \ (1,3,5) \leftrightarrow
    (2,2,4) \ or \ (0,4,4) \\
 && \leftrightarrow 
    (1,3,3) \leftrightarrow
    (2,2,2) .
\end{eqnarray*}
The metrics of the intersecting even-branes 
are the metrics of type IIA. 
The metrics of the intersecting odd-branes 
are the metrics of type IIB.

For four intersecting D-branes, we obtain the following metric:
\begin{eqnarray}
 (2,2,4,4) &=& \mbox{ {\scriptsize
        $\left\{ \begin{array}{c|cccccccccc}
     \x & \x & \x & -  & -  & -  & -  & -  & -  & - & \\
     \x & -  & -  & \x & \x & -  & -  & -  & -  & - & \\
     \x & \x & -  & \x & -  & \x & \x & -  & -  & - & \\
     \x & -  & \x & -  & \x & \x & \x & -  & -  & - & .
               \end{array} \right. $} } 
\end{eqnarray}
from $(2,2,5,5)_M$.
Applying the T-duality transformation, we can obtain the following metrics : 
\begin{eqnarray*}
 (3,3,3,3) \leftrightarrow (2,2,4,4) \leftrightarrow (1,3,3,5) 
 \leftrightarrow (0,4,4,4) .
\end{eqnarray*}

For two intersecting D-branes, we obtain the $(4,4)$ metrics: 
\begin{eqnarray*}
 (4,4) &=& \mbox{ {\scriptsize
        $\left\{ \begin{array}{c|cccccccccc}
      \x & \x & \x & \x & \x &  - &  - & -  & -  & - & \\
      \x & \x & \x &  - &  - & \x & \x & -  & -  & - & ,
        \end{array} \right. $} }   
\end{eqnarray*}
from $(5,5)_M$. Applying the T-duality transformation, 
we can obtain following metrics : 
\begin{eqnarray*}
 (4,4) \leftrightarrow (3,5) \leftrightarrow (2,6) , \\
 (3,3) \leftrightarrow (2,4) \leftrightarrow (1,5) , \\
 (2,2) \leftrightarrow (1,3) \leftrightarrow (0,4) ,
\end{eqnarray*}
in a non-intersecting direction, and 
\begin{eqnarray*}
 (4,4) \leftrightarrow (3,3) \leftrightarrow (2,2) ,
\end{eqnarray*}
in an intersecting direction.
We have assumed that these metrics have the three-dimensional 
transverse directions. Therefore all harmonic functions 
in these metrics only depend on $r = \sqrt{x^2+y^2+z^2}$.
Note that our notation is different from~\cite{BB}.

These metrics can also be obtained from the metrics
of three intersecting D-branes. For example, we find the following
relation:
\begin{eqnarray*}
 (3,3,5) \to (3,3) ,
\end{eqnarray*}
by assuming the third D5-brane's charge to vanish. 
For single D-brane, the T-duality transformation implies 
the following relation
\begin{eqnarray*}
 (6) \leftrightarrow (5) \leftrightarrow 
\cdots \leftrightarrow (0) .
\end{eqnarray*}

In order to obtain non-extremal metrics, we modify the above metrics as 
follows~\cite{CT} :

(1) We make the following replacements in the transverse 
space-time  part of the metric:
\begin{equation}
  dt^2  \to   f(r) dt^2 , \quad dr^2 \to f^{-1} (r) dr^2 ,
  \quad  f(r) = 1 - \frac{\mu}{r} .
\end{equation}
We also use the harmonic functions,
\begin{equation}
 H_i = 1 + \frac{{\Q}_i}{r} , \quad 
{\Q}_i= \mu\sinh^2 \delta_i ,
\end{equation}
for the constituent D2-branes, and 
\begin{equation}
 H_i = 1 + \frac{{\P}_i}{r} , \quad 
{\P}_i= \mu\sinh^2 \gamma_i ,
\end{equation}
for the constituent D4-branes.

(2) In the expression for the field strength ${\cal F}_4$ 
of the three-form field, we make the following  replacements: 
\begin{equation}
 H'_i \to H'_i = 1 + \frac{Q_i}{r + {\Q}_i - Q_i} 
  =\big[ 1- \frac{Q_i}{r} H_i^{-1} \big]^{-1} , \quad
 Q_i =\mu\sinh\delta_i\cosh\delta_i , 
\end{equation}
for the electric (D2-brane) part, and  
\begin{equation}
 H_i \to H'_i = 1 + \frac{P_i}{r} , \quad
 P_i =\mu\sinh\gamma_i\cosh\gamma_i\ , 
\end{equation}
for the magnetic (D4-brane) part.
Here $Q_i$ and $P_i$ 
are the electric and magnetic charges respectively.
The BPS limit is  realized when
$\mu \to 0 ,\  \delta_i\to \infty$, and $\gamma_i\to  \infty$,
while the charges $Q_i$ and $P_i$
are kept fixed. 
In this case ${\Q}_i=Q_i$ and  ${\P}_i =P_i$, so that  $H'_i=H_i$.
We note that the metric of two intersecting 
D-branes are not 
extremal in the BPS limit.

(3) In the case when the solution has a null isometry, 
i.e. the intersecting D-branes have a common string along 
some direction $y$, one can add momentum along 
$y$ by applying the coordinate transformation 
\begin{equation}
t'= \cosh \sigma \ t - \sinh \sigma \ y , \quad 
y'= - \sinh \sigma \ t + \cosh \sigma \ y , \label{sigma}
\end{equation}
to the non-extremal background which is obtained 
according to the above two steps. Then we obtain that 
\begin{eqnarray*}
 - f(r) dt^2 + dy^2\ &\to& - f(r) dt'^2 + dy'^2 \\
  &=& - dt^2 + dy^2 +
  \frac{\mu}{r} (\cosh \sigma \ dt - \sinh \sigma \ dy)^2, 
\end{eqnarray*}
where $\sigma$ is a boost parameter.
This transformation is called the boost transformation. 

\section{Surface Gravity and Compactification}
In ten-dimensions~\cite{CT,Tsey,GKT}, 
the metrics have the compactification radii. 
The radii are usually taken to be arbitrary. 
However for the generic radii, 
we find that the Euler numbers of the metrics are non-integers.
In this section, we consider the surface gravities 
of the compactified dimensions. 
We constrain the compactification radii 
using the surface gravity.
As a result, we obtain the integer Euler numbers. 

We consider a way to constrain the radii 
in the compactified directions.
We first consider the radius 
in the Euclidean time direction. 
We treat the Euclidean time coordinate 
as the polar angle in the $t-r$ directions.
We define the deficit angle $\delta$ 
in the $t-r$ directions as,
\begin{eqnarray*}
 \frac{1}{2}\int_0^\beta dt \int_{r_H}^\infty dr
  \sqrt{g_{tt}g_{rr}} R_{(2)} = 2\pi - \delta, 
\end{eqnarray*}
where $R_{(2)}$ is the scalar curvature 
in the $t-r$ directions. 
$\beta$ is the period of the time coordinate. 
$r_H$ is the radius of the event horizon, 
which satisfies that 
 $ g_{tt}\bigg|_{r = r_H} = 0 $.
The metric in the $t-r$ directions is 
\begin{eqnarray*}
 ds_{(2)}^2 = g_{tt} dt^2 + g_{rr} dr^2. 
\end{eqnarray*}
The scalar curvature $R_{(2)}$ and the extrinsic curvature $K$ 
are 
\begin{eqnarray*}
 R_{(2)} = - \frac{1}{\sqrt{g_{tt}g_{rr}}}
 \partial_r\bigg[\frac{1}{2}\frac{\partial_r g_{tt}}
 {\sqrt{g_{tt}g_{rr}}}\bigg], \quad
 K = \frac{1}{\sqrt{g_{rr}}}  		    
    \bigg[\frac{1}{2}\frac{\partial_rg_{tt}}{g_{tt}}\bigg] . 
\end{eqnarray*}
We define the Euler number $\chi$ 
from the Gauss-Bonnet theorem is 
\begin{eqnarray}
 2\pi \chi &=& \frac{1}{2}\int_0^\beta dt 
  \int_{r_H}^\infty dr \sqrt{g_{tt}g_{rr}}R_{(2)} 
  - \int_0^{\beta} dt \sqrt{g_{tt}}K 
     \bigg|_{r=\infty} \nonumber \\
   &=& \frac{\beta}{2}\frac{\partial_r g_{tt}}
 {\sqrt{g_{tt}g_{rr}}} \bigg|_{r=r_H} . \label{chi}  
\end{eqnarray}
If no singular effect of the event horizon exists, 
then the topology 
in $t-r$ directions is ${\bf R^2}$, 
then the Euler number is 1. 
Therefore, from (\ref{chi}) the radius $\beta$ 
in the time direction is
\begin{eqnarray*}
\beta = \frac{2\pi}{\kappa_t}, \quad 
  \kappa_t &\equiv& \frac{1}{2}\frac{\partial _r g_{tt}}
               {\sqrt{g_{rr}g_{tt}}}\bigg|_{r = r_H}, 
\end{eqnarray*}
where $\kappa$ is the surface gravity. 
$r_H$ is a radius of the black hole event horizon. 
We obtain that the deficit angle is $0$, 
and the Euler number is 1.
Then if we use the radius $\beta$ 
in the time direction, 
the singular effect of the event horizon $r_H$ is absent. 
Therefore we take this radius 
in the time direction 
to obtain the smooth geometry 
around the event horizon. 

For example, we find that the Euler numbers 
of the Schwarzschild metric, the Kerr-Newman metric, 
and the $U(1)$ dilaton metric to be integers 
by using the relation between 
the surface gravity and the compactification radius
(the inverse temperature). 


We extend this idea to the ten-dimensional manifolds
in order to obtain the integer Euler numbers. 
We next consider the compactification radii 
of the black holes in ten-dimensions. 
We treat the black holes which have six compactified 
directions. 
These compactified directions 
are 
the same as the Euclidean time direction, 
then the topology is ${\bf R^8} \times {\bf S^2}$.
The metric $g_{ii}$ of the compactified directions  
does not depend the variables 
of the compactified directions. 
Therefore we can choose 
one of the compactified directions 
and consider the Euler numbers 
in the $i-r$ directions partly. 

We consider the compactification radii in the $i-r$ 
directions, 
where $i$ is one of the compactified directions.
The topologies of the $i-r$ directions are ${\bf R^2}$. 
We define the deficit angle $\delta_i$ 
in the $i-r$ directions as 
\begin{eqnarray*}
 \frac{1}{2} \int_0^{\beta_i} dx^i \int_{r_H}^\infty dr 
  \sqrt{g_{ii}g_{rr}} R_{(2)} = 2\pi - \delta_i, 
\end{eqnarray*}
where $R_{(2)}$ is the scalar curvature 
in the $i-r$ directions. 
$\beta_i$ are the periods of the $i$-th coordinate. 
The metric in the $i-r$ directions is 
\begin{eqnarray*}
 ds_{(2)}^2 = g_{ii} dt^2 + g_{rr} dr^2. 
\end{eqnarray*}
The scalar curvature $R_{(2)}$ and the extrinsic curvature $K$ 
are 
\begin{eqnarray*}
 R_{(2)} = - \frac{1}{\sqrt{g_{ii}g_{rr}}}
 \partial_r\bigg[\frac{1}{2}\frac{\partial_r g_{ii}}
 {\sqrt{g_{ii}g_{rr}}}\bigg], \quad 
 K = \frac{1}{\sqrt{g_{rr}}}
    \bigg[\frac{1}{2}\frac{\partial_rg_{ii}}{g_{ii}}\bigg] .
\end{eqnarray*}
We define the Euler number $\chi$
from the Gauss-Bonnet theorem as 
\begin{eqnarray}
 2\pi \chi &=& \frac{1}{2}\int_0^{\beta_i} dx^i 
  \int_{r_H}^\infty dr \sqrt{g_{ii}g_{rr}}R_{(2)} 
  - \int_0^{\beta_i} dx^i \sqrt{g_{ii}}K 
     \bigg|_{r=\infty} \nonumber \\
 &=& \frac{\beta}{2}\frac{\partial_r g_{ii}}
 {\sqrt{g_{ii}g_{rr}}} \bigg|_{r = r_H}  . \label{chi2} 
\end{eqnarray}
If no singular effect of the event horizon exists, 
then the topology 
in $i-r$ directions is ${\bf R^2}$, 
and the Euler number is 1. 
Therefore, from (\ref{chi2})
we suppose that the compactification radius 
$\beta_i$ in the $i$-th direction is defined 
by the following "surface gravity $\kappa_i$": 
\begin{eqnarray*}
  \beta_i(r_H) &=& \frac{2\pi}{\kappa_i}, \quad 
  \kappa_{i}(r_H) 
   \equiv \frac{1}{2}\frac{\partial _r g_{ii}}
    {\sqrt{g_{rr}g_{ii}}}\bigg|_{r = r_H}, \\
  && (i=1 \cdots 6) .
\end{eqnarray*}
We obtain that the vanishing deficit angles 
in the compactified directions.
Therefore we use this radius $\beta_i$ 
in the compactified directions 
to 
obtain the smooth geometry 
around the event horizon. 

We obtain the integer Euler numbers 
for the ten-dimensional manifolds 
using these compactification radii.
For example, let us consider the following metric:
\begin{eqnarray*}
 (4,4) &=& \mbox{ {\scriptsize
        $\left\{ \begin{array}{c|cccccccccc}
      \x & \x & \x & \x & \x &  - &  - & -  & -  & - &  \\
      \x & \x & \x &  - &  - & \x & \x & -  & -  & - & .
                 \end{array} \right. $} }   
\end{eqnarray*}
In the non-extremal version of this metric, the event horizon is
located at 
\begin{eqnarray*}
 r_H = \mu .
\end{eqnarray*}
In the third direction, the compactification radius is 
\begin{eqnarray*}
 \beta_{3}(r_H) 
 = 4 \pi (\mu + \m_1)(\mu + \m_2)^{3/2}/(\m_1-\m_2)\sqrt{\mu} . \\
\end{eqnarray*}
Using this definition, we indeed find that the Euler number 
of this example to be an integer as follows:
\begin{eqnarray*}
 \chi &=& \frac{1}{(4\pi)^{10} 10!} \int dx^{10}  
 \epsilon^{abcdefghij}R_{ab}R_{cd}R_{ef}R_{gh}R_{ij} \\ 
      &=& \frac{1}{(4\pi)^{10}} 
          \int_{r_H}^{\infty} dr \int_0^{\beta_t(r)} dt  
          \int_0^{2\pi} d\phi \int_0^{\pi} d\theta \sin \theta \\
      &&  \times \prod_{i=1}^{6} \int_0^{\beta_i(r)} dx_i 
           R_{tr}R_{\theta\phi}R_{12}R_{34}R_{56} \\
      &=&  2 .
\end{eqnarray*}
We further obtain the same Euler number for all the other metrics
which are related to this metric by T-duality as it will be explained in the
next section. 
We emphasize again that the Euler numbers are non-integers
if we consider the compactification radii as arbitrary constants. 

\section{Black Hole Entropies and T-duality}
In this section, we study the entropies 
of the intersecting D-branes 
using the compactification radii 
as defined in the previous section. 
The entropies are calculated semiclassically by using $S=A/4G$
in this section. 
Here $A$ is the area of the event horizon, 
and $G$ is the Newton's constant. 
We obtain the entropies which are T-duality invariant. 
In the BPS limits ($\mu \to 0 $), 
we obtain the finite and non-vanishing entropies 
only for two intersecting D-branes. 
They are found to be proportional to the product 
of the charges of each D-brane. 
 We further study the entropies of boosted metrics. 
We obtain the relations 
between the entropies and the internal momenta of 
the intersecting D-branes. These relations are the same as 
that of the microscopic D-brane picture~\cite{SV,HS,MS}. 

We first study the entropies of the metric with the boost parameter 
$\sigma = 0$ 
in (\ref{sigma}). 
We define the proper length in the $i$-th direction as:
\begin{eqnarray*}
  L_i(r) &\equiv& 
\bigg|\int_{0}^{\beta_{i}(r)}\sqrt{g_{ii}}dx_i \bigg| 
 = 4\pi \bigg| \frac{\sqrt{g_{rr}}}{\partial_r (\ln g_{ii})} 
\bigg| , \\
  && ( i= 1,\cdots, 6 ) .
\end{eqnarray*}
Using these definitions, the entropies of the black holes are 
\begin{eqnarray}
 S = A_8/4G_{10} \bigg|_{r=\mu} 
   = L_1L_2L_3L_4L_5L_6A_{\theta\phi}/4G_{10} \bigg|_{r=\mu}. 
\end{eqnarray}
Here $A_8$ is the area of the event horizon, 
$A_{\theta\phi}$ is 
the area in the $\theta-\phi$ directions. 

For example, we consider the following metric : 
\begin{eqnarray*}
 (4,4) &=& \mbox{ {\scriptsize
        $\left\{ \begin{array}{c|cccccccccc}
     \x & \x & \x & \x & \x &  - &  - & -  & -  & - & \\
     \x & \x & \x &  - &  - & \x & \x & -  & -  & - & .
                 \end{array} \right. $} }   
\end{eqnarray*}
In the third direction, the proper length is 
\begin{eqnarray*}
 L_3 = \int_{0}^{\beta_{3}}\sqrt{g_{33}}dx_3 
     = 4 \pi (r + \m _1)^{5/4} (r + \m _2)^{5/4}
     /(\m _1-\m _2)\sqrt{r}, 
\end{eqnarray*}
where $\m_i$ are $\Q_i$ or $\P_i$. 
With this definition, we find the following relations :
\begin{eqnarray}
 L_3 = L_4 = L_5 = L_6 
 &=& 4\pi(r + \m_1)^{5/4} (r + \m_2)^{5/4}/(\m_1-\m_2)\sqrt{r} 
 \nonumber \\
 &\equiv& L_a, 
 \nonumber \\
 L_1 = L_2 
 &=& 4 \pi (r + \m_1)^{1/4} (r + \m_2)^{1/4}\sqrt{r} 
 \nonumber \\
 &\equiv& L_b. \label{pre}
\end{eqnarray}
Let us consider the proper lengths of the metrics 
which are dual to each other. 
If we apply the T-duality transformation to 
the fourth-direction of (4,4) type metric,
we obtain the (3,5) type metric, 
\begin{eqnarray*}
 (3,5) &=& \mbox{ {\scriptsize
        $\left\{ \begin{array}{c|cccccccccc}
     \x & \x & \x & \x & -  &  - &  - & -  & -  & - & \\
     \x & \x & \x &  - & \x & \x & \x & -  & -  & - & .
        \end{array} \right. $} }   
\end{eqnarray*}
For this metric, we also obtain the proper lengths 
which are the same with (\ref{pre}).
Then the entropies of (3,5) type metrics 
are the same with (4,4) type metrics. 
We also obtain the identical proper lengths 
for all the other metrics 
which are dual to each other. 
Therefore the entropies of the T-dual metrics are the same. 

We can also obtain the analogous relations with the different 
numbers of the D-branes. 
The area in the $\theta-\phi$ directions is 
\begin{eqnarray*}
 A_{\theta\phi} 
 = \int_{0}^{\pi} d\theta \int_{0}^{2\pi} d\phi \sin\theta 
   \sqrt{g_{\theta\theta}g_{\phi\phi}} 
 = 4 \pi (r + \m_1)^{1/2} (r + \m_2)^{1/2}r. 
\end{eqnarray*}
Then we find that the entropy of two intersecting D-branes is given by
\begin{eqnarray*}
S &=& A_8/4G_{10} \bigg|_{r=\mu} 
   = L_a^4L_b^2A_{\theta\phi}/4 \bigg|_{r=\mu} \\
  &=& (4\pi)^7(\mu+\m_1)^8(\mu+\m_2)^8 \\
  && /\bigg[(\m_1-\m_2)^4[\mu(\m_1+\m_2)+2\m_1\m_2]^2 \bigg] ,
\end{eqnarray*}
where $G_{10}$ is the ten-dimensional Newton's constant. 
For three intersecting D-branes, the entropy is  given by
\begin{eqnarray*}
S &=& A_8/4G_{10} \bigg|_{r=\mu} \\
   &=& (4\pi)^7((\mu+\m_1)^8(\mu+\m_2)^8(\mu+\m_3)^8\mu^2 \\
    && /\bigg[[\mu^2(\m_1+\m_2-\m_3) 
        + 2\mu \m_1\m_2 +\m_1\m_2\m_3] \\
    && \times [\mu^2(\m_2+\m_3-\m_1) 
        + 2\mu \m_2\m_3 +\m_1\m_2\m_3] \\
    && \times [\mu^2(\m_3+\m_1-\m_2) 
        + 2\mu \m_3\m_1 +\m_1\m_2\m_3] \bigg]^2 .
\end{eqnarray*}
For four intersecting D-branes, the entropy is  
\begin{eqnarray*}
S &=& A_8/4G_{10} \bigg|_{r=\mu} \\
  &=& (4\pi)^7((\mu+\m_1)^8(\mu+\m_2)^8(\mu+\m_3)^8(\mu+\m_4)^8 \\
   && /\mu^6\bigg[
        [\mu^2(\m_1+\m_2-\m_3-\m_4) +2\mu(\m_1\m_2-\m_3\m_4) \\
   &&   \ + (\m_1+\m_2)\m_3\m_4-(\m_3+\m_4)\m_1\m_2] \\
   &&   \times 
        [\mu^2(\m_1+\m_3-\m_2-\m_4) +2\mu(\m_1\m_3-\m_2\m_4) \\
   &&   \ + (\m_1+\m_3)\m_2\m_4-(\m_2+\m_4)\m_1\m_3] \\
   &&   \times
        [\mu^2(\m_1+\m_4-\m_2-\m_3) +2\mu(\m_1\m_4-\m_2\m_3) \\
   &&   \ + (\m_1+\m_4)\m_2\m_3-(\m_2+\m_3)\m_1\m_4] \bigg] ^2 . \\
\end{eqnarray*}
We also list the entropy formula for single brane:
\begin{eqnarray*}
 S = A_8/4G_{10} \bigg|_{r=\mu} = (4\pi)^7\mu^6(\mu+\m_1)^8/\m_1^6 .
\end{eqnarray*}

If we consider the BPS limit (namely $\mu \to 0$), 
we can obtain the finite and non-vanishing entropies 
only for the metrics of two intersecting D-branes. 
In order to interpret these entropies,
we define the quantized D-brane charges 
in the wrapping directions 
except for the intersecting directions. 
They are the integer numbers. 
In the case of $(4,4)$ case, we define the  
quantized  magnetic charges of the D4-branes as 
\begin{eqnarray*}
 N_1 &=& \frac{L_a^2}{4\pi} \int {\cal F}_{\theta\phi12} 
     d\Omega 
 \bigg|_{r=1} \\
     &=& (4\pi)^2\m_1(\m_1\m_2)^{5/2}/(\m_1-\m_2)^2, \\ 
 N_2 &=& \frac{L_a^2}{4\pi} \int {\cal F}_{\theta\phi12} 
     d\Omega 
 \bigg|_{r=1} \\
    &=& (4\pi)^2\m_2(\m_1\m_2)^{5/2}/(\m_1-\m_2)^2 .
\end{eqnarray*}
Using these charges, we find the entropies as 
\begin{eqnarray}
 S = (4\pi)^3 N_1 N_2 /4 . \label{44}
\end{eqnarray}
In the T-dual cases of this metric, (3,5) and (2,6), we define
\begin{eqnarray*}
 N = \left\{ 
 \begin{array}{ll}
  \frac{L_a^2}{4\pi}\int *{\cal F} & \mbox{(electric)}, \\
  \frac{L_a^2}{4\pi}\int {\cal F}  & \mbox{(magnetic)}, 
 \end{array} \right. 
\end{eqnarray*}
where $\ast$ is Hodge dual in four-dimensions. 
Then we find that the entropies for all metrics 
of two intersecting D-branes as the same with (\ref{44}).
Therefore we observe that the entropies of 
two intersecting D-branes 
are the product of the quantized 
charges of each D-brane. 

Next we consider the case of the large boost parameter 
($\sigma \gg 1$). 
In this case, the metric becomes 
\begin{eqnarray*}
 - f(r) dt^2 + dy^2\ &\to& - f(r) dt'^2 + dy'^2 \\
   &=& - dt^2 + dy^2 
   + \frac{\mu}{r} (\cosh \sigma \ dt - \sinh \sigma \ dy)^2 \\
   &\sim& - dt^2 + dy^2 
       + \frac{\mu\cosh^2 \sigma}{r} (dt-dy)^2 .
\end{eqnarray*}
Here we have introduced $\mu' = \mu \cosh^2 \sigma$. 
Then the event horizon is at $r_H= \mu'$. 
The proper length in the $y$- direction is 
\begin{eqnarray*}
 L_y = 4\pi[(\m_1+r)(\m_2+r)]^{5/4} \sqrt{r}
     /(2r^2-2\m_1\m_2) . 
\end{eqnarray*}
For $(3,3), (2,4)$ and $(1,5)$ cases, we find the following entropy
formula as 
\begin{eqnarray*}
 S = L_a^4L_bL_yA_{\theta\phi}/4\bigg|_{r=\mu'}.
\end{eqnarray*}
In the $(4,4), (3,5)$ and  $(2,6)$ cases, we have two isometric directions. 
If we consider the case that only one of these directions is 
boosted, the entropy is given by 
\begin{eqnarray*}
 S = L_a^4L_bL_yA_{\theta\phi}/4\bigg|_{r=\mu'}.
\end{eqnarray*}
We may further consider the case that 
the metric is boosted in two different directions.
Let us assume that the $y_2$ direction is boosted first 
and then the $y_1$ direction is boosted. 
After such a process, we obtain the following metric as  
\begin{eqnarray*}
 - f(r) dt^2 + dy_1^2 + dy_2^2 \ 
   &\to& - f(r) dt'^2 + dy_1'^2 + dy_2'^2 \\
   &=& - dt^2 + dy_1^2 + dy_2^2 \\
    && + \frac{\mu}{r} 
   (\cosh^2 \sigma_1\cosh^2 \sigma_2\ dt 
    - \sinh \sigma_1\ dy_1 
    - \cosh^2 \sigma_1\sinh^2 \sigma_2 dy_2)^2 \\
   &\sim& - dt^2 + dy_1^2 +dy_2^2
    + \frac{\mu\cosh^2 \sigma_1 \cosh^2 \sigma_2}{r} 
    (dt-dy_2)^2. 
\end{eqnarray*}
Therefore we obtain the same metric 
with the boost in the single direction. 
As a result, we find that the entropies 
of the boosted metrics are the same. 

According to the definitions in~\cite{SV,HS,MS},
we define the quantized internal momentum $p$ 
and the charge $n$ in the direction of $y$. 
Here $n$ is a number of the state level of the effective
conformal field theory in the D-brane picture. 
The momentum $p$ is quantized in term of 
the surface area of the event horizon $A_8$.  
We define the charge and the momentum as 
\begin{eqnarray*}
   n &=& L_y p ,\\
   n &=& \int dA_8 \mu'/r \bigg|_{r=\mu'}\\
     &=& L_a^4L_bL_yA_{\theta\phi} \\
     &\sim& (\m_1\m_2)^6 /(\m_1-\m_2)^4. 
\end{eqnarray*}
Using these quantities, we obtain the entropies in the BPS limit 
with $\sigma \gg 1$ as follows, 
\begin{eqnarray*}
 S \sim \sqrt{N_1N_2n}. 
\end{eqnarray*}
This entropy formula of the boosted metrics is the same with 
that of the microscopic 
D-brane picture~\cite{SV,HS,MS}. 

\section{Conclusion}
We have studied the Euler numbers and the entropies 
of the non-extremal black holes 
which are constructed 
by the intersecting D-branes in ten-dimensions. 
The Euler numbers are generally integers.  
In ten-dimensions, the metrics have 
the arbitrary compactification radii. 
However for the generic radii, 
the Euler numbers of the metrics are non-integers.
To avoid these difficulties, 
we need the way to constrain 
the compactification radii. 

The another physical reasoning that we need to fix 
the compactification radii 
is in order to avoid the singular effects 
of the horizon in the compactified directions. 
We have discussed this point in the section 3 .
We consider the way to define the period in the time  directions.
If no singular effects of the horizon exist in $t-r$ directions, 
then the topology of these directions are ${\bf R^2}$, 
and the Euler number of these directions is 2. 
Using the Gauss-Bonnet theorem,
we find that 
the Euclidean time coordinate 
have the period which is 
the inverse of the surface gravity. 
Therefore we need to take this period 
in the time direction 
to avoid the singular effects of the horizon.
Similarly, we need to fix 
the compactification radii to avoid the singular effects 
of the horizon in the compactified directions.
We find the necessity that 
the compactification radii are the inverses of 
the surface gravity in the compactified directions 
using the Gauss-Bonnet theorem. 

We identify the inverse of the surface gravities as 
the compactification radii 
in order to obtain the integer Euler numbers. 
The entropies of the black holes are 
T-duality invariant when we use 
the compactification radii 
which are defined by surface gravities. 
In the BPS limit, we have the finite and non-zero entropies 
only with two intersecting D-branes. 
We have introduced the quantized D-brane 
charges $N_1$ and $N_2$. 
We find the common relation
of these entropies as $S \sim N_1N_2$, 
when the boost parameter vanishes. 
Therefore we observe that the entropies 
are proportional to the product of the charges of each D-brane.

In the case of the large boost parameter, 
we also obtain the entropy formula for 
the black holes constructed by two intersecting D-branes. 
The entropies of the  boosted black holes 
are also T-duality invariant.
We have introduced the quantum number $n$. 
Here $n$ is a number of the state level of the 
effective conformal field theory in the D-brane picture. 
The entropies are $S \sim \sqrt{N_1N_2n}$. 
The entropies of the boosted black holes can be interpreted 
as the entropies of the microscopic states 
in the D-brane picture. 

We need to consider the microscopic interpretation of  
the entropies in association with the non-boosted 
black holes in the D-brane picture.
The entropies of the non-boosted black holes reduce 
to $S \sim N_1N_2$. 
The black holes have no momenta $n$ in this limit.
On the other hand, 
the entropies in the microscopic 
D-brane picture $S \sim \sqrt{N_1N_2n}$
are obtained for the large momenta.
Therefore the entropy formula for the non-boosted black holes 
could be entirely different from that for the boosted black holes
which agrees with
the microscopic D-brane picture.
The microscopic entropies in the D-brane picture 
with small or zero momenta 
were not discussed so far. 
We further need to study the microscopic 
interpretation of the entropies 
with the momenta $n \sim 0$ in the D-brane picture. 

We have proposed a method to constrain the compactification 
radii of the non-extremal black holes of 
the intersecting D-branes. 
We have identified the inverse of the surface gravities as 
the compactification radii. 
We have correctly obtained 
the integer Euler numbers of the black holes. 
Although we do not know another way to constrain 
the compactification radii,
we have not shown that the way 
to constrain the compactification radii
is unique. 
Therefore another way to constrain 
the compactification radii might be found. 
We further need to study the way 
and the physical reasoning 
to constrain the  compactification radii. 

\acknowledgements 
We thank Y. Kitazawa for discussions and 
for carefully reading the manuscript 
and suggesting various improvements.



\end{document}